\documentclass[aps,pra, twocolumn,showpacs]{revtex4}
\usepackage{amsmath}
\usepackage{epsfig}
\usepackage{amssymb}
\usepackage{dcolumn}
\usepackage{graphicx}
\usepackage{dcolumn}

\begin{document}
\date{\today}

\title{Stabilization of high-order solutions of the cubic Nonlinear Schr\"odinger Equation}

\author{Adrian Alexandrescu} 
\email{adrian.alexandrescu@uclm.es}
  
\author{Gaspar D. Montesinos}
 
\author{V\'{\i}ctor  M. P\'erez-Garc\'{\i}a}
\email{victor.perezgarcia@uclm.es}
\affiliation{Departamento de Matem\'aticas, E. T. S.  Ingenieros Industriales and \\ Instituto de Matem\'atica Aplicada a la Ciencia y la Ingenier\'{\i}a (IMACI) \\ Avda. Camilo Jos\'e Cela, 3, Universidad de Castilla-La Mancha, 13071 Ciudad Real, Spain}

\begin{abstract}
In this paper we consider the stabilization of non-fundamental unstable stationary solutions of the cubic nonlinear Schr\"odinger equation. 
Specifically we study the stabilization of radially symmetric solutions with nodes and asymmetric complex stationary solutions. For the first ones we find partial stabilization similar to that recently 
found for vortex solutions while for the later ones stabilization does not seem possible.
\end{abstract}


\maketitle

\section{Introduction}

Nonlinear Schr\"odinger Equations (NLS) are one of the most important models of mathematical physics arising
 in a great array of contexts \cite{Sulem,Vazquez} as for example in semiconductor 
electronics \cite{Soler}, optics in nonlinear media \cite{Kivshar}, photonics \cite{Hasegawa}, plasmas \cite{Dodd}, 
fundamentation of quantum mechanics \cite{fundamentals}, dynamics of 
accelerators \cite{Fedele}, mean-field theory of Bose-Einstein condensates \cite{Dalfovo} or in 
biomolecule dynamics \cite{Davidov}, to cite a few examples.

It is well known that in multidimensional scenarios there may appear  concentration phenomena and collapse 
depending on the initial configuration \cite{Sulem}. Let us write the model equation in the form 
\begin{equation}
i u_z = -\frac{1}{2} \Delta u + g(z) |u|^2 u,
\label{NLS}
\end{equation}
on $\mathbb{R}^2$ with $\Delta = \partial^2/\partial x^2 + \partial^2/\partial y^2$ and initial data $u_0(x,y)$. It is well-known that, when $g(z) = g_0<0$, the solutions of Eq. \eqref{NLS} with $L^2\left(\mathbb{R}^2\right)$ norm
\begin{equation} \label{norm}
N(u) \equiv \| u\|_2^2 = \int_{\mathbb{R}^2} |u|^2,
\end{equation}
 such  that $g_0\| u\|^2$ is larger than a critical value $N_0$ may undergo collapse (i.e. catastrophic formation of very sharp 
gradients and concentration phenomena for $u$). However, when $g_0 \| u\|^2 < N_0$ there cannot be collapse \cite{Sulem}. 

In the context of the analysis of the propagation of Kerr beams in layered optical media it was first explored the possibility that making the  
nonlinear coefficient $g(z)$ to oscillate with the independent variable between values corresponding to the collapsing and expansion regimes could lead to 
collapse supresion \cite{Berge}.

This idea was later explored \cite{Malomed} in the same context and exported to the field of Bose-Einstein condensation \cite{Ueda,Abdullaev,Gaspar}. 
The stabilized structure was identified as a pulsating Townes soliton after some rearrangement of initial data \cite{Gaspar,IMACS} that is able to propagate without essential distortions for very large distances. 
Some rigorous results concerning early time collapse (i.e. the situation in which collapse cannot be avoided) where found in \cite{Konotop}.

It has been only recently that the theoretical concept of stabilized solitons has been demonstrated in the laboratory in Optical experiments  \cite{Kevre,Kevre2}.

 Stabilized solitons have also been studied in three-dimensional scenarios \cite{Gaspar,Adhi1,Ueda2}. 
Also, they have been considered in vector media leading to the so-called stabilized vector solitons 
\cite{PROLO,Chaos}. Finally, the possibility of existence of stabilized solitons with different nonlinearities and dimensionalities has been studied in 
Ref. \cite{CQ}.

The idea of stabilized solitons and nonlinearity management has also inspired some related mathematical research, with a focus on the averaged and 
collapse-preventing properties \cite{Peli,Peli2,Peli3}. However, a rigorous theoretical mathematical description of the stabilization process is still missing \cite{teoro}.

The reviews \cite{Fatkh,Malo2} offer a panoramic vision of the field of stabilized solitons.

In this paper we want to complement the present knowledge on stabilized solitons with a numerical study of the stabilization of more complex stationary 
solutions of the nonlinear Schr\" odinger equation. Although the Townes soliton can be stabilized, vortices have been found more difficult to stabilize 
\cite{PRE} since the periodic modulation of the non-linear coefficient alone is not enough to achieve stabilization.

 The aim of this work is to investigate the possibility of stabilizing different types of stationary solutions of the NLS equation. Specifically, we will study the
stabilization of higher order radially symmetric solutions in scalar and vector media to check if the simpler structure of these solutions with respect to 
vortices allows them to be stabilized. We will also study the stabilization of more complex asymmetric solutions of the NLSE described in 
Ref. \cite{Alfimov}.
 
The paper is organized as follows: in Sec. \ref{one-component} we study the stabilization of higher-order radially symmetric solutions in scalar layered 
media and find that it is not posible to achieve stabilization of these structures in this simple way. In Sec. \ref{two-component} we show the stabilization 
of these structures by the addition of a second stabilizing component, i.e. in the vector case. In Sec. \ref{sec_asym}  we consider the possibility of stabilizing 
higher order asymmetric solutions.  Finally, in Sec. \ref{conclu} we summarize our conclusions.

\section{Stabilization of higher-order radially symmetric solutions in scalar layered media}
\label{one-component}

First we start by analyzing the simple case of scalar fields whose propagation is governed by the NLS equation (\ref{NLS}), with $g(z)=g_0$ constant.  In this case, the equation for stationary solutions $u(z,x,y) = \phi(x,y) e^{i\lambda z}$, is 
\begin{equation}\label{estacionaria}
-\lambda \phi = -\frac{1}{2} \Delta \phi + g_0 |\phi|^2 \phi.
\end{equation}
It is well known \cite{Sulem}, that Eq. \eqref{estacionaria} has an infinite number of solutions with radial symmetry each having a finite number of nodes. We will denote these 
solutions by $R_0, R_1, R_2, ...$ labelling them by their number of nodes. Their norms satisfy $N_0 < N_1 < N_2 ... $. $R_0$ is the ground state, also called the Townes soliton, which plays an important role in the theory 
since $N_0$ sets a critical value for the norm below which collapse cannot exist. Due to the scaling symmetry of the nonlinear Schr\"odinger equation, these solutions exist for any value of $\lambda$.

\begin{figure}[htb]
\epsfig{file=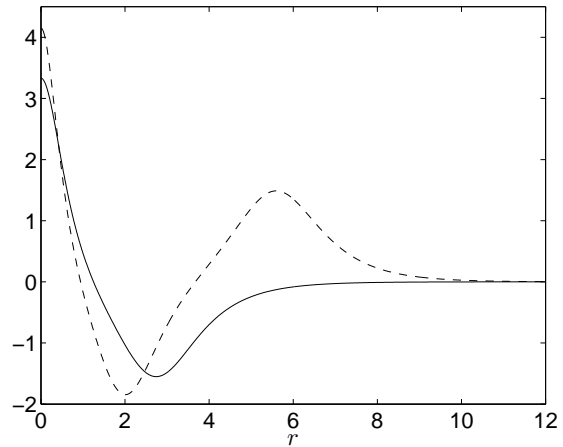, width=\columnwidth}
\caption{Plot of $R_1$  (solid line) and $R_2$ (dashed line), normalized to unity, for  $g_0=-0.5, \lambda=0.5$.  For these solutions 
$N_{1}\simeq 77.17$ and $N_{2}\simeq 195.84$, respectively.}
\label{higher_order_solutions}
\end{figure}

The first two higher order stationary solutions with radial 
symmetry, i.e. $R_1,$ and $R_2$ calculated by using a standard shooting method, 
 are depicted in Fig. \ref{higher_order_solutions}.  All of our radially symmetric solutions to be presented in this paper have been computed by this method and then 
 interpolated onto a two-dimensional rectangular grid. After this injection, we have used a Newton relaxation method in order to increase the accuracy of the computed stationary solutions.

\begin{figure}[htb]
\epsfig{file=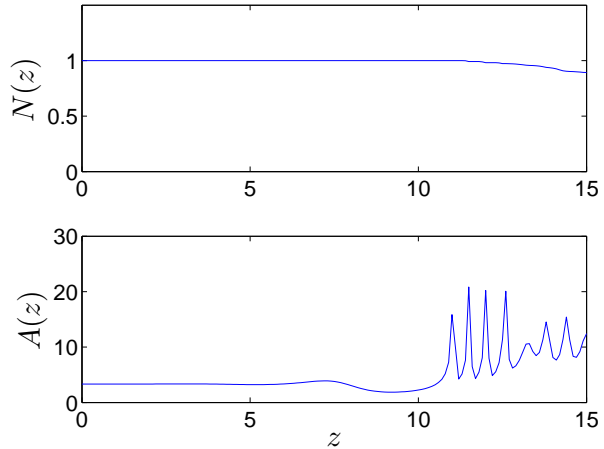, width=\columnwidth}
\caption{Propagation of the norm given by Eq. (\ref{norm}) and maximum amplitude $A(z) = \max_{(x,y)} |u(x,y,z)|$ for the solution of Eq. (\ref{NLS}) with initial data $u(x,y,0) = R_1(r)$  through an homogeneous nonlinear media, i.e. $g(z)=-0.5$.}
\label{firstnode_free_evolution}
\end{figure}

We have checked the ``stationarity" of our numerically found solutions by propagating them through an homogeneous nonlinear medium subject to the perturbation coming from 
the numerical errors (both on the initial data and due to the roundoff error during the evolution). All our numerical simulations to be presented in this paper have been done using a second order in time and spectral in space 
split-step Fourier algorithm with absorbing boundary conditions to get rid of the outgoing radiation.
In Fig. \ref{firstnode_free_evolution} we present the propagation of the norm and amplitude taking $u(x,y,0) = R_1(r)$. As the stationary solutions of Eq. (\ref{estacionaria}) are all unstable in the context of Eq. (\ref{NLS}), one expects that sooner or later the instability will set in. In the case of the Townes 
and vortex solitons the instability sets in by $z\simeq 15$ and $z\simeq 3$, respectively  \cite{PRE}. From the propagation of the norm and amplitude 
shown in Fig. \ref{firstnode_free_evolution}, we can guess that, in our case, a collapse destroying the structure of this stationary state occurs near $z\simeq 11$.

 We have tried to apply the stabilization technique based on the modulation of the nonlinear coefficient  to  the first radially excited solution by setting $g(z)=g_0+g_1\cos(\Omega z)$, 
 with $g_0=-0.5$, $g_1=-1.5$ and $\Omega=100$  (the parameters were set according to criteria established in Ref. \cite{Gaspar,IMACS}) and taking $u(x,y,0)  =R_1(r)$. This mechanism allows a 
 stable propagation of the Townes soliton over long distances of more than 
  400 units in adimensional units \cite{Gaspar} but the same  mechanism enhances  only  slightly the stability of singly-charged vortex solitons \cite{PRE}. 
 In principle our stationary solution has a simpler structure than the vortex and one could expect better results, but we will see that this is not true.
  
\begin{figure}[htb]
\epsfig{file=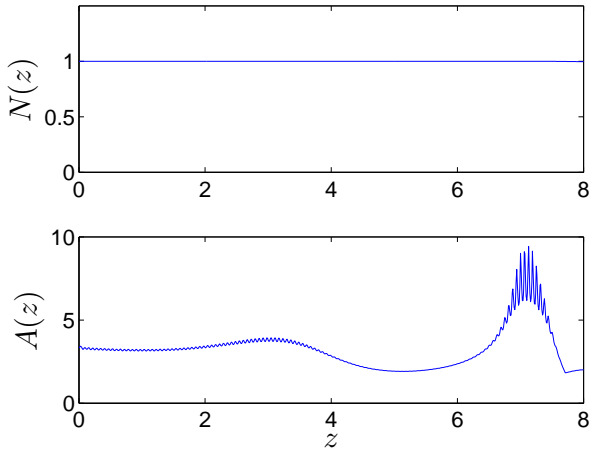, width=\columnwidth}
\caption{Propagation of the norm given by Eq. (\ref{norm}) and maximum amplitude $A(z) = \max_{(x,y)} |u(x,y,z)|$ for the solution of Eq. (\ref{NLS}) with initial data $u(x,y,0) = R_1(r)$ under the 
effect of a periodic modulation of the nonlinear coefficient 
$g(z)=g_0+g_1\cos(\Omega z)$ where $g_0=-0.5$, $g_1=-1.5$ and $\Omega=100$.}
\label{firstnode_gmod_evolution}
\end{figure}

  Our results are shown in Fig. \ref{firstnode_gmod_evolution}. We can see how the inclusion of this periodic modulation leads to a shorter stable propagation distance of the structure and that collapse occurs by $z\simeq 6.5$

We have tried to stabilize this configuration by choosing different parameters for the modulation with similar results. We think that the reduction of the lifetime of the unstable stationary structure can be understood  
by considering the exchange of energy, i.e. energy flow, between the substructures of the first excited radially symmetric stationary solution: the central peak and its 
surrounding ring. During the defocusing ($g(z)>0$) stages both substructures spread and overlap in the region where the field was zero previously. Then, in the focusing 
($g(z)<0$) stage the energy from the overlapping region flows mainly to the central peak. Therefore, with each focusing step the central peak is supplied with more and more energy, and this process 
leads finally to appearance of collapse (see the sharp amplitude peaks for $z\simeq 7$ in Fig. \ref{firstnode_gmod_evolution}).


\section{Stabilization of higher-order radially symmetric solutions in vector layered media}
\label{two-component}

\subsection{Motivation and model}

Our next idea is to try to stabilize the excited radially symmetric stationary 
solutionss by using a stabilized Townes soliton as a guide in which the higher order solution could be stabilized as it was done in Ref. \cite{PRE} for vortices. Hence, we shift our attention to vector systems described by the following set of coupled NLS equations:
\begin{subequations}
\begin{eqnarray}
            i\frac{\partial u_1}{\partial z}  =  -\frac{1}{2}\Delta u_1 +g(z)\left( a_{11}|u_1|^2 + a_{12}|u_2|^2\right)u_1 \label{nonlinear_equation_1},\\
            i\frac{\partial u_2}{\partial z} =  -\frac{1}{2}\Delta u_2 +g(z)\left( a_{21}|u_1|^2 + a_{22}|u_2|^2\right)u_2 \label{nonlinear_equation_2}.
\end{eqnarray}
\label{nonlinear_system_1}
\end{subequations}
Defining $\alpha=N(u_2)/N(u_1)$ and $\tilde u_i=u_i/N_i$, $i=1,2$ Eqs. (\ref{nonlinear_system_1}) become 
\begin{subequations}
\begin{eqnarray}
            i\frac{\partial \tilde u_1}{\partial z}  & =  -\frac{1}{2}\Delta \tilde u_1 +g(z)N_1\left( a_{11}|\tilde u_1|^2 + \alpha a_{12}|\tilde u_2|^2\right)\tilde u_1,  \label{scaled_non-linear_equation1}\\
            i\frac{\partial \tilde u_2}{\partial z}  & =  -\frac{1}{2}\Delta \tilde u_2 +g(z)N_1\left( a_{21}|\tilde u_1|^2 + \alpha a_{22}|\tilde u_2|^2\right) \tilde u_2 ,  \label{scaled_non-linear_equation2}
\end{eqnarray}
\label{nonlinear_system_2}
\end{subequations}
 For simplicity, we will discard the tilde in what follows and take both components to be normalized.
 
 The parameter $\alpha$ is a measure of the strength of the interaction between both components, which could be 
accomplished experimentally by launching beams of different energies.

\subsection{Case $\alpha=0$}
\label{linear_case}

In the limit $\alpha\rightarrow 0$, corresponding to the case when the norm of one of the components is much smaller than the other,  Eqs. (\ref{nonlinear_system_2}) become
\begin{subequations}
\label{towlin}
\begin{eqnarray}
i\frac{\partial u_1}{\partial z} & = & -\frac{1}{2}\Delta u_1 + g(z) N_1 a_{11}|u_1|^2 u_1,  \label{townes_equation}\\
i\frac{\partial u_2}{\partial z} & =&  \left[-\frac{1}{2}\Delta u_2 + g(z) N_1 a_{21}|u_1|^2\right] u_2. \label{linear_equation}
\end{eqnarray}
\end{subequations}
Eq. (\ref{townes_equation}) is a scalar NLS equation with a modulated nonlinear coefficient and thus admits solutions in the form of stabilized Townes solitons. Eq. (\ref{linear_equation}) is a
 linear Schr\"odinger equation for $u_2$ in a trapping potential generated by $u_1$.
First we look for stationary solutions of Eqs.  (\ref{towlin}) when $g(z) = g_0$ of the form
\begin{eqnarray}
u_1(\boldsymbol{r},z)  & = & \phi_{\lambda}(r)\exp\left(-i\lambda z\right), \\
  u_2(\boldsymbol{r},z) & = & \varphi_{\mu}(r)\exp(-i\mu z),
\end{eqnarray}
i.e. solutions of the nonlinear eigenvalue problem
\begin{subequations}
\label{towlin_sys}
\begin{eqnarray}
\lambda \phi & = & -\frac{1}{2}\Delta \phi + g_0 N_1 a_{11}|\phi|^2 \phi,  \label{townes_equation_sys}\\
\mu \varphi & =&  \left[-\frac{1}{2}\Delta \varphi + g_0 N_1 a_{21}|\phi|^2\right] \varphi. \label{linear_equation_sys}
\end{eqnarray}
\end{subequations}

Obviously, Eq. \eqref{townes_equation} is equivalent to Eq. (\ref{estacionaria}) and thus we get all of its stationary solutions, for instance, the Townes soliton.  We will look for stationary 
solutions of Eq. (\ref{linear_equation_sys})  with radial symmetry beyond the nodeless one $\varphi_0$. This implies that the effective potential 
\begin{equation}
V(r)=a_{21}g_0N_1|\phi_{\lambda}|^2,
\end{equation}
must support at least two bound states. The number of radially symmetric bound states supported by a two-dimensional potential
is bounded by the inequality \cite{newton}:
\begin{equation}
N_{2D,l=0} < 1+\frac{1}{2}
\frac{\int_{\mathbb{R}^2} d r\; d r^\prime \; r r^\prime \:V(r)  \;V(r^\prime)\; \left|\ln\left(\tfrac{r}{r^\prime}\right) \right|}{\int d r \:r V(r)}.
\label{upper_bound}
\end{equation}
This inequality sets a necessary condition to have the required bound state (i.e. that with one node) for fixed  $g_0=-2\pi$ by fixing a lower bound for the value of $a_{21}$ required to obtain $N_{2D,l=0}\geq 2$.
In Table \ref{lambda_table} we show the lower bounds  and the numerical results for the number of bound states obtained for several values of the coefficient $a_{21}$.
We can see that for  $a_{12}=4$ we can have two modes in the effective potential generated by $R_0$, the second one being a radially symmetric function with one node. 
\begin{table}[htb]
\begin{tabular}{ccc}
$a_{21}$&$N_{2D,l=0}$&Bound states eigenvalues \\
\hline
1 & $<1.39$ & $\lambda_0\simeq 0.575$ \\
2 & $<1.79$ & $\lambda_0\simeq 1.925$ \\
3 & $<2.19$ & $\lambda_0\simeq 3.558$ \\
4 & $<2.58$ & $\lambda_0\simeq 5.335$, $\lambda_1\simeq 0.195$\\
\hline
\end{tabular}
\caption{Upper bound set by Eq. (\ref{upper_bound}) on the number of bound states in two dimensions with radial symmetry and without angular momentum of the 
potential $V(r) = a_{21}g_0N_1|R_0|^2$. $\lambda_0$ and $\lambda_1$ are the approximate
eigenvalues of the ground and first excited states of $V(r)$, respectively, found numerically.}
\label{lambda_table}
\end{table}

\begin{figure}[htb]
\epsfig{file=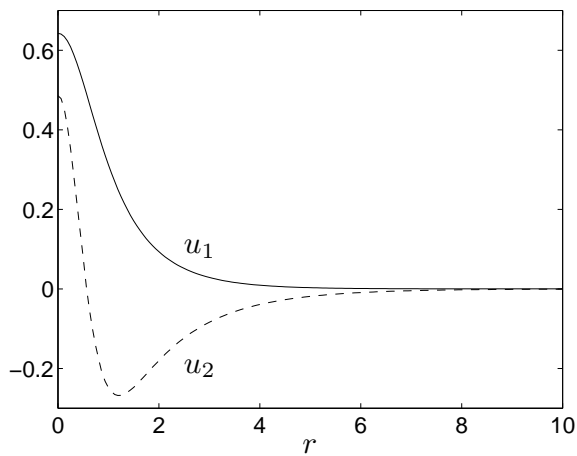, width=\columnwidth}
\caption{[Color online] Spatial radial profile (normalized to one and for $\lambda = 0.5$) of the stationary solution of Eqs. (\ref{towlin_sys}) of the form $\phi = R_0$.(solid line) and the first excited state with radial symmetry $\varphi = \varphi_1$ (dashed line).}
\label{townes_firstnode}
\end{figure}

We have computed the  profile of the radially symmetric solution of Eq. (\ref{linear_equation}) with one node in  $\varphi_1$ which exists for $a_{21} = 4$ by 
using a standard shooting method, which is shown in Fig. \ref{townes_firstnode} 

We have propagated numerically the initial data $u_2 = \varphi_1$ according to Eq. (\ref{linear_equation}) alone but now with a periodically varying nonlinear coefficient. 
During the propagation the potential $a_{21} g(z) |R_0|^2$ oscillates with frequency $\Omega$ due to the periodic modulation of coefficient
$g(z)$, hence, the potential will change its behavior periodically from attractive to repulsive. In general, for the linear situation described by 
Eq.  (\ref{linear_equation}), one may achieve a non-dispersive propagation 
for small oscillations of $g(z)$, i.e. without being necessary to change its sign. However, as we are interested to extend these 
results to the full nonlinear equations, we must construct $g(z)$ with an alternating sign because analytical results 
\cite{Konotop, Gaspar} and computer simulations \cite{Gaspar} have revealed that  this is a requirement to get stabilized solitons.

\begin{figure}[htb]
\epsfig{file=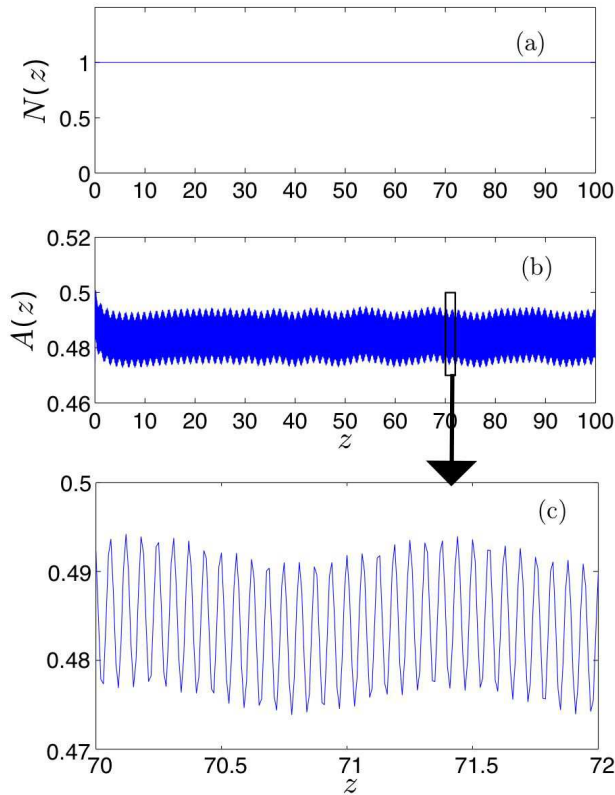, width=\columnwidth}
\caption{[Color online] Evolution of the  (a) norm $N(z)$ and (b) maximum value of the amplitude $A(z)$ described by the Eq. (\ref{linear_equation}) 
for parameter values $g_0=2\pi$, $g_1=8\pi$ and $\Omega=100$. (c) 
Detailed view of the amplitude  oscillations in the propagation range $z \in [70,72]$.}
\label{one-component_time_evolution}
\end{figure}

Using results from the quantum mechanical theory of fast perturbations \cite{galindo} we have chosen a set of parameters, $g_0=-2\pi$,
 $g_1=8\pi$ and $\Omega=100$ for which the oscillating behavior of $g(z)$ should maintain the profile of the excited solution in $u_2$.
Moreover, this set of parameters leads to an stabilized Townes soliton in component $u_1$ \cite{IMACS}. 
In Fig. \ref{one-component_time_evolution} we can see that although the maximum value 
of the amplitude exhibits oscillations related to the periodic modulation of $g(z)$, the norm remains constant which 
indicates that no radiation is emitted while the potential $a_{21} g(z) |u_1|^2$ is switched from attractive to repulsive.

\begin{figure}[htb]
\begin{center}
\epsfig{file=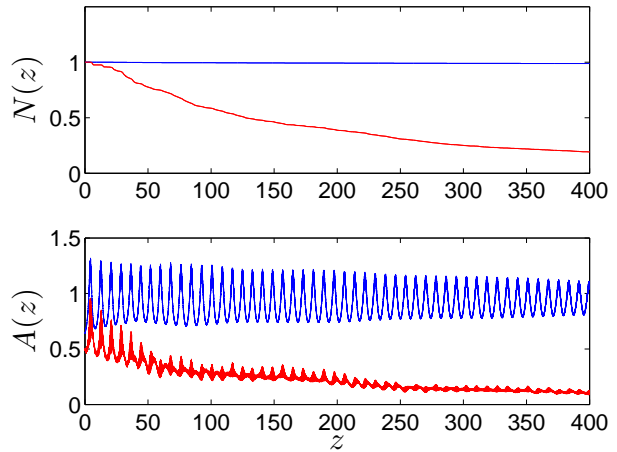, width=\columnwidth}\\
\caption{[Color online] Propagation of initial data of the form  $u_1 = R_0, u_2 = U_1$ under Eq. (\ref{towlin}) with $\alpha=0$, with $g_0=-2\pi$, $g_1=8\pi$ and $\Omega=100$. Shown are the norm $N(z)$ and
 maximum amplitude $A(z)$ for $u_1$ (blue) and $u_2$ (red). 
\label{two-component1_time_evolution}}
\end{center}
\end{figure}

Next we take $u_1 = R_0, u_2 = U_1$ as initial data for Eqs. (\ref{towlin}).
This situation is described by the case of $\alpha\simeq 0$, which means that the energy injected into the medium 
by the second component $u_2$ is much smaller in comparison with that of the first component $u_1$ \cite{PRE}.
 The dynamics of the most important parameters of the system is plotted in Figure \ref{two-component1_time_evolution}.

We want to remark that the time evolution of $|u_1|^2$  exhibits high and low frequency oscillations corresponding to the periodic modulation of 
$g(z)$ and to the internal dynamics of  Eq. \ref{townes_equation}, respectively \cite{Gaspar}. Therefore, the potential experienced by $u_2$ includes this oscillation pattern.
As the low frequency oscillations do not fulfill the requirements \cite{PRE} derived from the theory of fast perturbations, they lead to energy emission from 
the second component $u_2$. The radiation emission is linked to the decreasing value of the $u_2$ norm: as the outgoing radiation hits the boundary of 
the computational domain it is removed by an absorbing potential. Nevertheless, the first component evolves unperturbed as a stabilized Townes soliton
\cite{Gaspar}.
\begin{figure}[htb]
\begin{center}
\epsfig{file=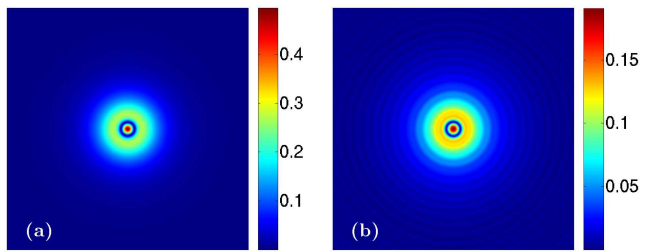, width=\columnwidth}
\caption{[Color online] Pseudocolor plot of the intensity $|u_2(x,y,z)|^2$ for the same simulation as in Fig. \ref{two-component1_time_evolution} on the spatial region $[-10,10]\times [-10,10]$ for (a) $z=0$ and (b) $z=400$.}
\label{two-component1_time_evolution2}
\end{center}
\end{figure}

The spatial intensity profile of the propagating beam $u_2$, see Fig. \ref{two-component1_time_evolution2},
has similar shape to the initial one (see Fig. \ref{higher_order_solutions}). However, the maximum of  the intensity distribution in 
Fig. \ref{two-component1_time_evolution}(b) is about three times smaller than the corresponding value of the initial one due to the energy loss during the propagation.

\subsection{Weak nonlinear coupling}
\label{nonlinear_case}

\begin{figure}[htb]
\begin{center}
\epsfig{file=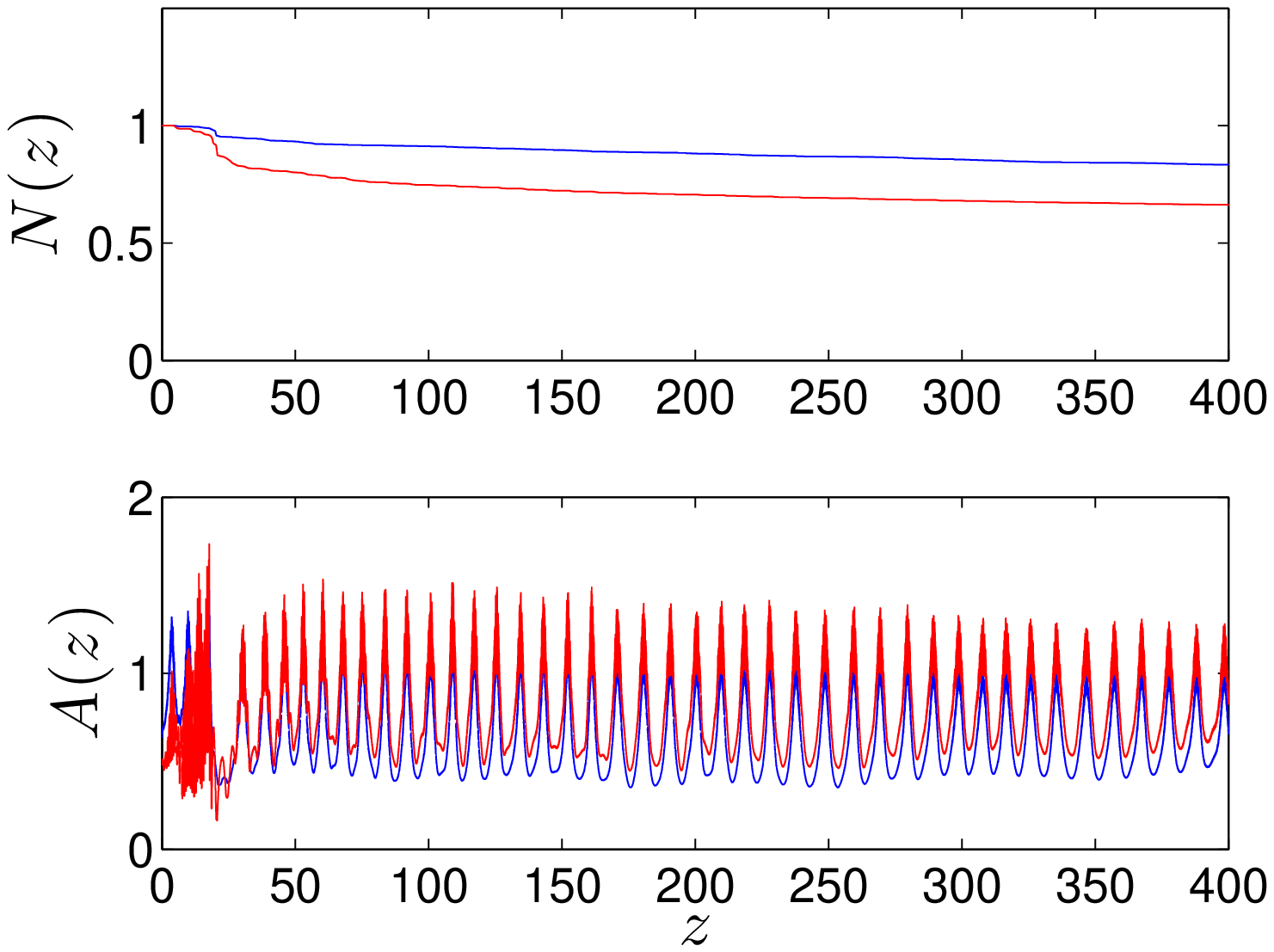, width=\columnwidth}\\
\caption{[Color online] Propagation of initial data of the form  $u_1=R_0, u_2 = U_1$ under Eq. (\ref{towlin}) with $\alpha=0.1$, with $g_0=-2\pi$, $g_1=8\pi$ and $\Omega=100$. Shownare the norm $N(z)$ and maximum amplitude $A(z)$ for $u_1$ (blue) and $u_2$ (red).
\label{two-component2_time_evolution}}
\end{center}
\end{figure}

Finally we turn on the nonlinear coupling in system (\ref{nonlinear_system_2}) by setting $\alpha=0.1$. 
The dynamics of the system parameters is presented in Figure \ref{two-component2_time_evolution}. 
\begin{figure}[htb]
\begin{center}
\epsfig{file=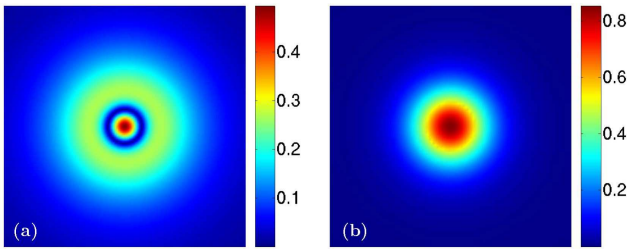, width=\columnwidth}
\caption{Pseudocolor plot of the intensity $|u_2(x,y,z)|^2$ for the same simulation as in Fig. \ref{two-component2_time_evolution} on the spatial region for (a) $z=0$ on the spatial region $[-4,4]\times[-4,4]$ and (b) $z=400$.
on the spatial region $[-2,2]\times[-2,2]$.}
\label{two-component2_time_evolution2}
\end{center}
\end{figure}
One can see that the propagation of the excited state is drastically changed in comparison with the previous analyzed cases. 
Under the effect of the nonlinear coupling, both components $u_1$ and $u_2$ reshape their transverse spatial profile by emitting energy. 
After a short propagation distance, $z=25$, the excited solution losses its spatial shape, acquiring a profile which
resembles to the Townes soliton and which will be propagate along with the Townes soliton. We may say that both initial profiles decay or
readjust (with energy loss) their shape, eventually leading to the formation of nodeless stabilized vector solitons \cite{PROLO}.
Nevertheless, for smaller values of $\alpha$, e.g. $\alpha=0.05$, we recover the behaviour described in the previous subsection, with shape preservation accompanied by energy emission.

\begin{figure}[htb]
\epsfig{file=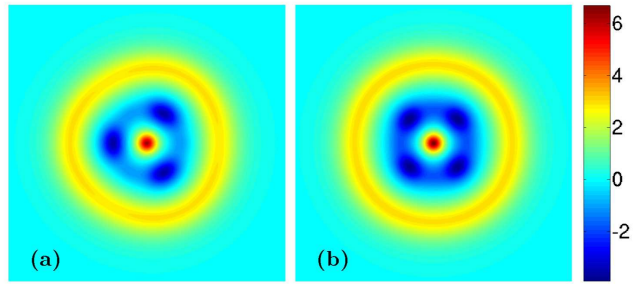, width=\columnwidth}
\caption{[Color online] Pseudocolor plots of asymmetric solutions of Eq. (\ref{estacionaria}) (a) $C_3$ invariant solution with norm $N_3\simeq 627.68$ and (b) $C_4$ invariant solution with norm $N_4\simeq 723.91$. The spatial region shown is  $(x,y) \in [-10,10]\times [-10,10]$.}
\label{solutions_alfimov}
\end{figure}

We think that the mechanism described at the end of Section \ref{one-component} is responsable for the decay of component $u_2$ when the nonlinear coupling is large enough.
The changes taking place in the component $u_2$ are then coupled back to component $u_1$, which leads to energy emission, as seen by the decreasing norm
of $u_1$ in Figure \ref{two-component2_time_evolution}.

 \begin{figure}[htb]
\epsfig{file=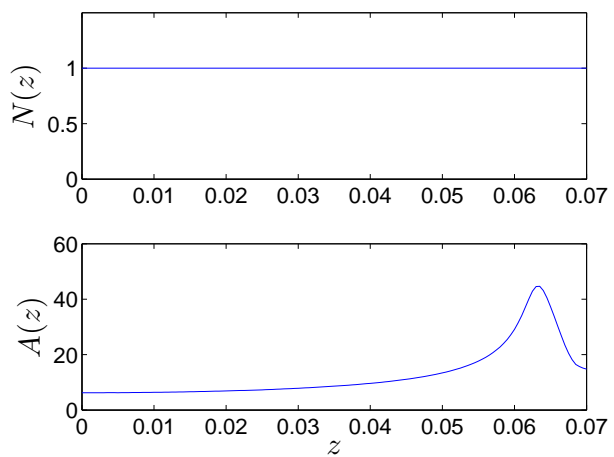, width=\columnwidth}
\caption{[Color online] Propagation through an homogeneous nonlinear medium of the norm $N(z)$ and maximum amplitude $A(z)$ for the stationary solution depicted in Fig. \ref{solutions_alfimov}(a).}
\label{nu3_free_evolution}
\end{figure}

\begin{figure}[htb]
\epsfig{file=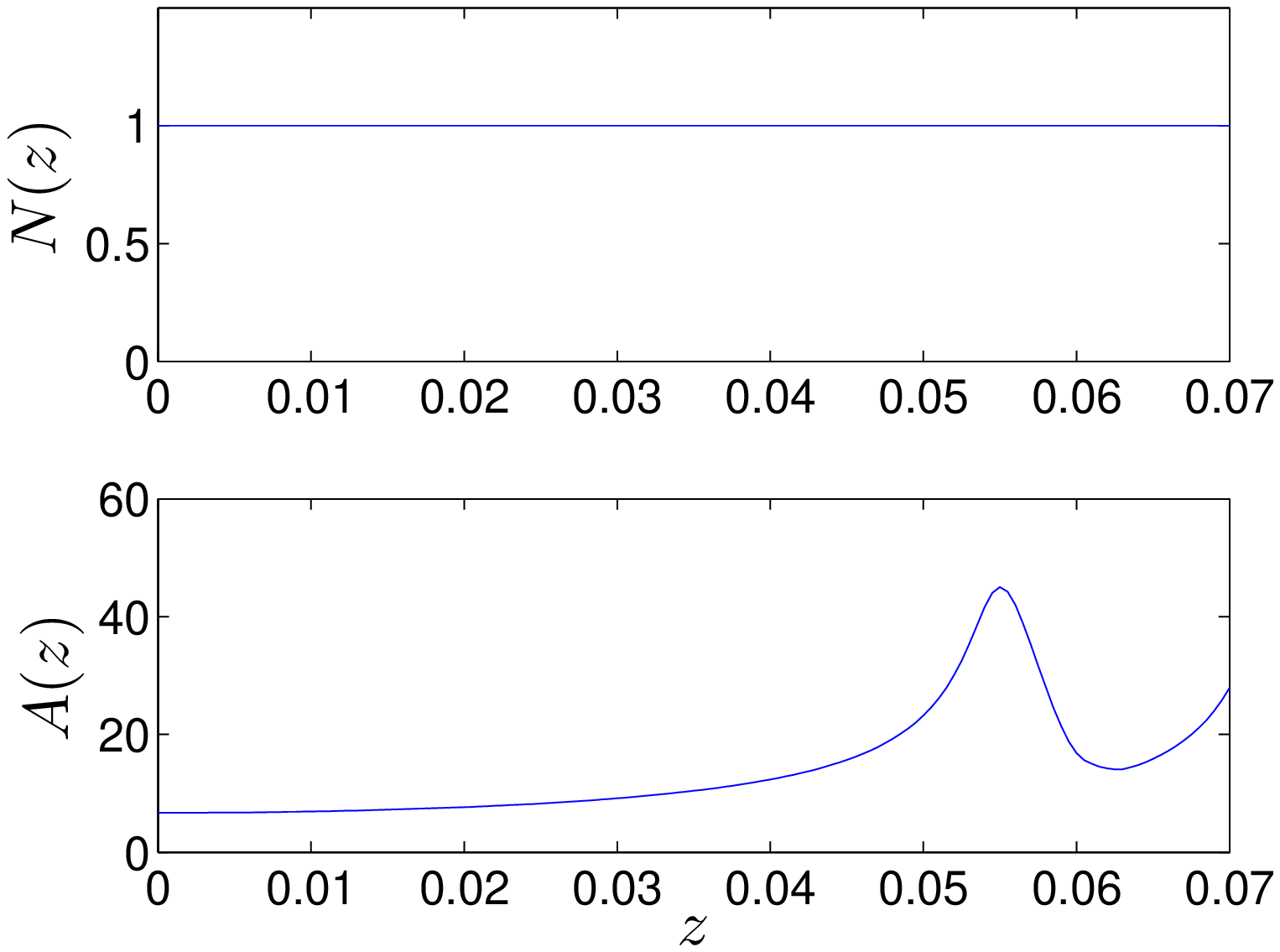, width=\columnwidth}
\caption{[Color online] Propagation through an homogeneous nonlinear medium of the norm $N(z)$ and maximum amplitude $A(z)$ for the stationary solution depicted in Fig. \ref{solutions_alfimov}(b).}
\label{nu4_free_evolution}
\end{figure}
\section{Solutions without radial symmetry}
\label{sec_asym}

Equation (\ref{estacionaria}) has more solutions beyond those having radial symmetry. For instance, 
 Alfimov and coworkers \cite{Alfimov}, using branching-off techniques from the theory of dynamical systems constructed solutions having nontrivial discrete rotational symmetries.
Two examples of those solutions are shown in Fig. \ref{solutions_alfimov}. In all of them we observe that a large central peak is surrounded by smaller ones having the prescribed 
discrete symmetry plus an outer ring.

Since these solutions are also unstable, their free propagation leads to collapse as shown in 
Figs. \ref{nu3_free_evolution} and \ref{nu4_free_evolution}. Since their norms are much above the critical one both solutions suffer a fast instability to collapse, the constant behavior of the
norm indicating the fact that outgoing radiation waves have no time to escape from this system before the instability develops. The sensitivity of those configurations to collapse manifests itself in the fact that
the instability distances are two orders of magnitude smaller than those of simpler solutions such as the Townes, vortex or first radially excited solitons.

We have tried to use the modulation of the nonlinear coefficient to stabilize the stationary solutions presented in Fig. \ref{solutions_alfimov}.  However, the time evolution of the beam parameters does not change appreciably and in particular, the emergence of collapse cannot be delayed. We have tried unsucessfully different sets of parameters to achieve stabilization. Intuitively, it seems difficult to be able to stabilize those complicated structures because of the coexistence of different types of substructures (peaks, rings, etc) and the fact that their norm ar many times the critical one (see the caption of Fig. \ref{solutions_alfimov}).

\section{Conclusions}
\label{conclu}

In this paper we have complemented previous knowledge on stabilized solitons of the Nonlinear Schr\"odinger Equation by studying
 numerically the possibility of stabilizing excited stationary solutions with radial symmetry in both (i) scalar and  (ii) vectorial layered media.  
The scalar system is unable to stabilize radially excited states due to the internal dynamics, i.e. energy flow, of this state which leads to shorter stable propagation distances of the  beam and collapse when compared with the
propagation of the same beam through homogeneous media. In the case of vector layered media we have shown how weakly coupled beams when one of the components is chosen to be the 
Townes soliton and the other an unstable radially excited solution can be stabilized. In this situation it is found that the radially excited state radiates continuously energy
while preserving an intensity shape similar to the initial profile. 

In both scalar and vectorial layered media, the stabilization of solutions bearing nontrivial structure is difficult to achieve. This fact is due to the ireversibility of the internal energy flows 
between the wave substructures, e.g. rings, peaks, during propagation.
Additional stabilizing mechanisms, like spatially inhomogeneous nonlinearities depending on the transverse variables, might help in stabilizing states with nontrivial structures as it happens in the case of non-oscillating structures \cite{U1,U2,U3}.

Finally, we have tried unsuccesfully to stabilize complex asymmetric solutions of Eq. (\ref{NLS}). The complex structure of these solutions and the fact that their power is many times the critical power makes stabilization probably impossible.
\acknowledgements

We want to thank G. Alfimov for discussions. This work has been partially supported by grants FIS2006-04190 (Ministerio de Educaci\'on y Ciencia, Spain) and PAI-05-001 (Consejer\'{\i}a de Educaci\'on y Ciencia de la Junta de Comunidades de Castilla-La Mancha, Spain),

\end{document}